\documentclass[twocolumn]{aastex61}

\newcommand\aastex{AAS\TeX}

\usepackage{amsmath}

\received{ }
\revised{ }
\accepted{ }
\submitjournal{ApJ}

\shorttitle{\aastex\ sample article}
\shortauthors{Pascucci et al.}

\begin{document}

\title{A Universal Break in the Planet-to-Star Mass-Ratio Function of {\it Kepler} MKG stars}

\correspondingauthor{Ilaria Pascucci}
\email{pascucci@lpl.arizona.edu}

\author{Ilaria Pascucci}
\affiliation{Lunar and Planetary Laboratory, The University of Arizona, Tucson, AZ 85721, USA}
\affiliation{Earths in Other Solar Systems Team, NASA Nexus for Exoplanet System Science}

\author{Gijs D. Mulders}
\affiliation{Lunar and Planetary Laboratory, The University of Arizona, Tucson, AZ 85721, USA}
\affiliation{Earths in Other Solar Systems Team, NASA Nexus for Exoplanet System Science}

\author{Andrew Gould}
\affiliation{Department of Astronomy, Ohio State University, 140 W. 18th Ave., Columbus, OH 43210, USA}

\author{Rachel Fernandes}
\affiliation{Lunar and Planetary Laboratory, The University of Arizona, Tucson, AZ 85721, USA}
\affiliation{Earths in Other Solar Systems Team, NASA Nexus for Exoplanet System Science}



\begin{abstract} 
We follow the microlensing approach and quantify the occurrence of {\it Kepler} exoplanets as a function of planet-to-star mass ratio, $q$, rather than planet radius or mass. For planets with radii $\sim1-6$\,R$_\oplus$ and periods $<$ 100 days, we find that, except for a normalization factor, the occurrence  rate vs $q$ can be described by the same broken power law with a break at $\sim 3 \times 10^{-5}$ independent of host type for hosts below 1\,M$_\odot$. These findings indicate that the planet-to-star mass ratio is a more fundamental quantity in planet formation than planet mass.
We then compare our results to those from microlensing for which the overwhelming majority satisfies the  $M_{\rm host}<1$\,M$_\odot$ criterion. The break in $q$ for the microlensing planet population, which mostly probes the region outside the snowline,  is $\sim$3-10 times higher than that inferred from {\it Kepler}. Thus, the most common planet  inside the snowline is $\sim$3-10 times less massive than the one outside. With rocky planets interior to gaseous planets, the Solar System broadly follows the combined mass-ratio function inferred from {\it Kepler} and microlensing. However,  the exoplanet population has a less extreme radial distribution of planetary masses than the Solar System. Establishing whether the mass-ratio function beyond the snowline is also host type independent will be crucial to build a comprehensive theory of planet formation.
\end{abstract}

\keywords{methods: data analysis $-$ planets and satellites: formation $-$ planetary systems}



\section{Introduction} \label{sec:intro}
The past decade has seen an exponential increase in the number of discovered exoplanets (e.g., \citealt{Fischer2014}). Yet, our knowledge of the exoplanet population is rather sparse, especially at a few au from the star. This radial distance is of particular interest as it is close to the so-called snowline in a protoplanetary disk (e.g., \citealt{Mulders2015a}), the location  beyond which water vapor condenses onto ice. At the snowline the surface density in solids increases by a factor of a few to several (e.g., \citealt{Min2011}), thus promoting the formation of larger planetary cores while the gaseous disk is still present.  In addition, the isolation mass for protoplanets, as well as for pebble accretion, increases with radial distance from the star (e.g., \citealt{KI2002,Lambrechts2014}).
As a consequence, a zeroth order expectation of in-situ planet formation models is that planets inside the snowline have lower masses than those formed beyond, an expectation that is fulfilled in the Solar System.

The {\it Kepler} survey is the most comprehensive survey for exoplanets inside the snowline (e.g., \citealt{Borucki2017}). The planet orbital period and planet-to-star radius are the parameters that are best determined by the transiting technique used by {\it Kepler}. Among the many interesting discoveries, {\it Kepler} has identified a large population of Super-Earths and many compact multi-planet systems (e.g., \citealt{WF2015}). It has also established that small planets ($< 3$\,R$_\oplus$) are more abundant around smaller M dwarf stars (e.g., \citealt{Mulders2015b}).

Microlensing is currently the most sensitive technique to a range of planetary masses, well below Neptune's mass (e.g., \citealt{Beaulieu2006}), close to and beyond the snowline (e.g., \citealt{GL1992}) .
The planet-to-star mass ratio (hereafter, $q$) is well determined with this technique while the host-star mass often remains unknown. Recently, \citet{Suzuki2016} found  that the exoplanet occurrence rate vs $q$ can be well described by a broken power law with a break at $\sim 10^{-4}$, which corresponds to $\sim$20\,M$_\oplus$ for the median host-star mass of $\sim 0.6$\,M$_\sun$. \citet{Udalski2018} employed a new method to re-analyze the systems with $q< 10^{-4}$, refined the power law index in this regime, and confirmed that it is different from the one at higher $q$. 

Here, we apply a "microlensing point of view" to the {\it Kepler} exoplanets by quantifying their occurrence rate vs $q$ rather than planet mass or radius.
We find that the occurrence rate has a break at $\sim 3 \times 10^{-5}$ independent of host type for hosts below 1\,M$_\odot$ (Section~\ref{sec:kepler}), which points to an almost universal mass-ratio function. 
Next, we compare our results to those from microlensing (Section~\ref{sec:kepler-moa}) for which the overwhelming majority satisfies the  $M_{\rm host}<1$\,M$_\odot$ criterion. We find that the break in $q$ is $\sim$3-10 times higher for microlensing planets, which are located outside the snowline, compared to {\it Kepler} planets, which are inside. We discuss the implications of both of these results in Section~\ref{sec:discussion}.

\section{The {\it Kepler} mass-ratio function} \label{sec:kepler}
We start from the new DR25 {\it Kepler} catalog \citep{Thompson2017} and select dwarfs following \citet{Huber2016}, see their Equation (9). First, we compute planet occurrence rates as a function of planet radius and per spectral type as in \citet{Mulders2015b} with updated detection efficiencies calculated using KeplerPORTS \citep{Burke2015}. Only
 the most reliable planet candidates are included in our analysis, those with Robovetter score $> 0.9$. Then, we convert planet radii into planet masses using the best-fit mass-radius relation by \citet{CK2017}. As in the microlensing studies, we choose a spacing in planet mass that is constant in base-10 logarithm (hereafter, log) and compute $q$ by dividing the planet mass at the center of each bin by the median stellar mass in each spectral type group (0.42, 0.73, 0.91, 1.1\,M$_\sun$ for M, K, G, F dwarfs respectively).  As uncertainty we take the planet occurrence rate divided by the square root of the number of planet candidates per mass bin.

\begin{figure}[ht!]
\plotone{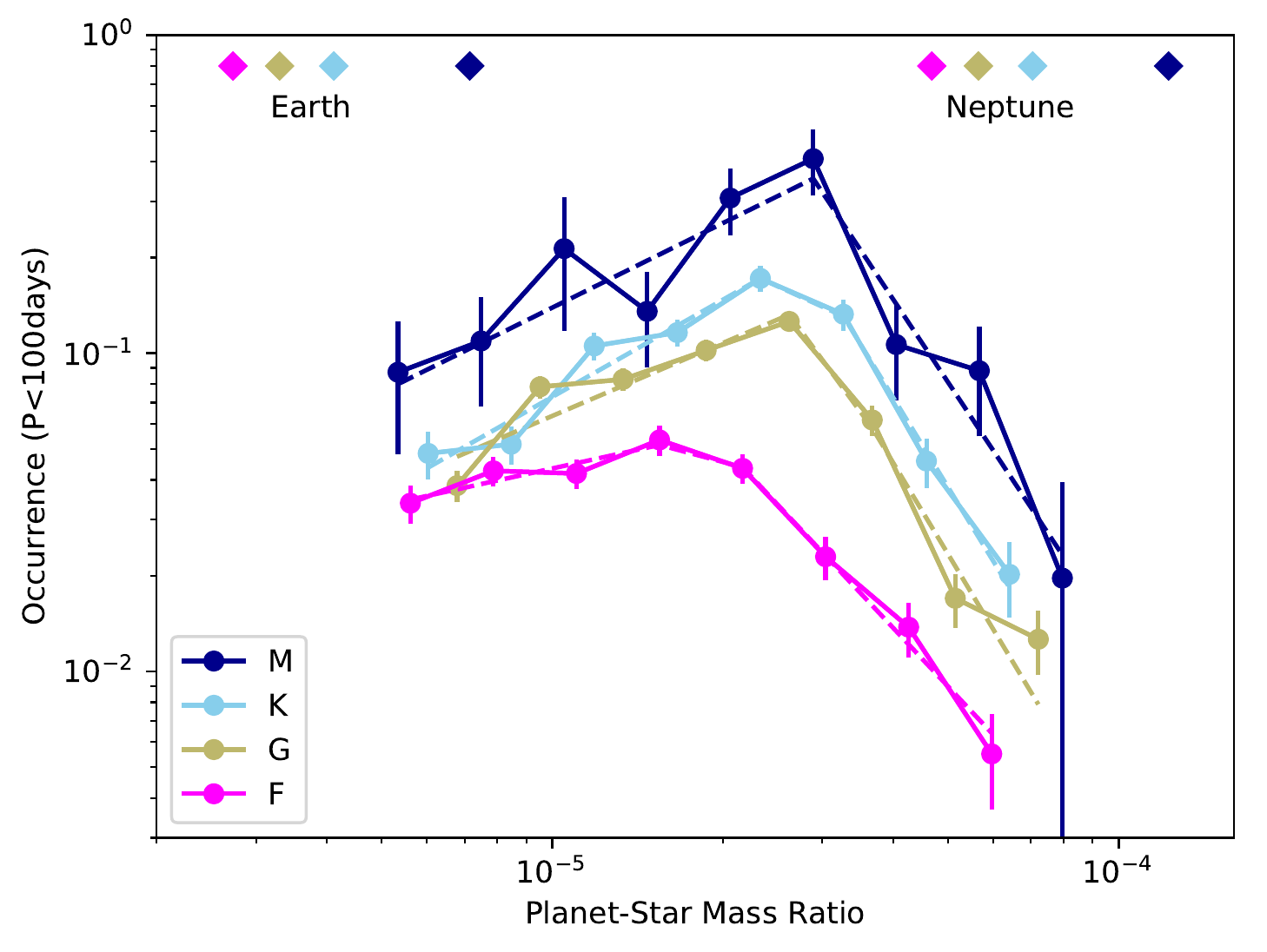}
\caption{Planet occurrence rate as a function of planet-to-star mass ratio ($q$), for spectral types M-F. Dashed lines indicate the best broken power law fits to the data. 
Colored diamonds at the top of the figure show the corresponding  $q$ for an Earth- and a Neptune-mass planet. 
\label{fig:kepler}}
\end{figure}

Figure~\ref{fig:kepler} shows the occurrence rate vs $q$ for planets with a period $<$100\,days\footnote{There are very few exoplanet candidates around M dwarfs with a period longer than 100\,days. Moreover, this period excludes the region beyond the snowline even for the median {\it Kepler} M star (e.g., \citealt{Mulders2015a}). } color-coded by stellar spectral type. We limit the plot and following analysis to $q \sim (0.5-8) \times 10^{-5}$, planet radii $\sim 1-6$\,R$_\oplus$, due to incompleteness in the {\it Kepler} survey at the lower end and degeneracy between planet mass and radius at the upper end. Both issues are mitigated when using the forward modeling approach employed by EPOS (see Section~\ref{sec:kepler-moa}).

First, we note that the shape of the distributions is fairly similar and the peak appears to occur at the same $q$ for MKG hosts. To compare the distributions in a more quantitative way, we use the {\tt SciPy} implementation of the non-parametric Anderson-Darling test. The null hypothesis that the MKG samples come from the same parent distribution can only be rejected at a significance level of 20\% while adding the F sample reduces the level to 4\%. This indicates that the occurrence rate of F hosts is more different than that of MKG hosts.

Next, we fit the occurrence rates with a broken power law. Except for a normalization factor $A$, the mass-ratio function is identical for hosts with $M_{\rm host}<1$\,M$_\odot$
\begin{equation} 
{dN\over d\log q} = A \biggl( {q\over q_{\rm br}} \biggr)^n
\label{eq:broken}
\end{equation}
with $q_{\rm br} \sim 2.8 \times 10^{-5}$, $n \sim 1 \, (q<q_{\rm br})$, and $n \sim -2.9 \, (q>q_{\rm br})$, see Table~\ref{tab:bestfit} for the best fit parameters\footnote{We have also employed the CRAN package {\tt segmented} and found breaks in the distributions that are the same, within the uncertainties, to the $q_{\rm br}$ reported in  Table~\ref{tab:bestfit}.}. This suggests an almost universal mass-ratio function and implies that the most common planetary mass  inside the snowline is not fixed but rather increases linearly with stellar mass, from $\sim$3.5-4.5\,M$_\oplus$ around M dwarfs up to $\sim$8-9\,M$_\oplus$ around G  stars. 

\citet{Suzuki2016} already commented on the similarity of literature-based  {\it Kepler} occurrence rates for M- and GK-dwarfs as a function of planet mass. Without fitting the curves, they noted peaks at $\sim$6\,M$_\oplus$ and $\sim$8\,M$_\oplus$,  respectively.  We found that the difference with our values is partly due to different {\it Kepler} samples\footnote{The M and GK-dwarf occurrence rates in \citet{Suzuki2016} are not from the same {\it Kepler} data release.} but also by their use of a mass-radius relation that imposes a fixed transition at 1.6\,R$_\oplus$. This relation alters the shape of what {\it Kepler} actually measures, which is the occurrence vs planet radius.

\begin{deluxetable}{l cccc}[ht!]
\tablecaption{{\it Kepler} Mass-Ratio Function Parameters\label{tab:bestfit}}
\tablewidth{0pt}
\tablehead{
\colhead{Parameter} &
\colhead{M host} & \colhead{K host} & \colhead{G host} & \colhead{F host}
}
\startdata
$A (\times 10^{-1})$         & 3.5$\pm$0.9& 2.1$\pm$0.2 & 1.37$\pm$0.09 & 0.56$\pm$0.06\\
$q_{\rm br} (\times 10^{-5})$ &  2.9$\pm$0.3 & 2.8$\pm$0.1      & 2.8$\pm$0.2 & 1.9$\pm$0.1 \\
$n (q<q_{\rm br})$        & 0.9$\pm$0.3    & 1.0$\pm$0.1      & 0.76$\pm$0.08 & 0.4$\pm$0.2\\
$n (q>q_{\rm br})$       & -2.7$\pm$0.7    &  -2.9$\pm$0.4    & -2.9$\pm$0.4 & -1.9$\pm$0.2\\
\enddata
\end{deluxetable}

\section{{\it Kepler} vs Microlensing} \label{sec:kepler-moa}
In this section we use the Exoplanet Population Observation Simulator (EPOS, Mulders et al. in prep.) to better account for the incompleteness of the {\it Kepler} survey at small planet radii and the dispersion in the mass-radius relation. Given that the shape of the occurrence vs $q$  is the same for MKG stars (Section~\ref{sec:kepler}) and G stars are $\sim$65\% of this sample, we run EPOS on this spectral type as representative for the low-mass stars. When comparing the occurrence at specific $q$ we will take into account the different normalization factors summarized in Table~\ref{tab:bestfit}.

We adopt a broken power law similar to Equation~(\ref{eq:broken}) but in 2D to simultaneously describe the planet occurrence rate vs orbital period ($P$) and vs planet mass ($M$), see Appendix~\ref{app:epos} for details. The planet radius, which is the quantity measured by {\it Kepler}, is calculated using the probabilistic mass-radius relation by \citet{CK2017}. A synthetic planet population is generated by randomly drawing a planet period and mass from the 2D broken power law and by assigning the planet randomly to one of the stars monitored by {\it Kepler}. Assuming an isotropic distribution of system orientations, EPOS then calculates a transiting planet population. Finally, the transiting population is subject to the {\it Kepler} detection efficiency from the latest DR25 data release and compared to the observed planet population within the region of interest, periods $< 100$\,days and radii $1-6$\,R$_\oplus$.

To estimate the best fit parameters of the broken power law and their confidence intervals, EPOS uses a Markov Chain Monte Carlo approach with the {\tt emcee} Python algorithm \citep{Foreman-Mackey2013} and uniform priors.
The broken power laws provide a satisfactory fit to the marginalized period and radius distribution, see Figure~8 in Mulders et al. in prep.  Corner plots showing the projections of the likelihood function are presented in Appendix~\ref{app:epos}. 

EPOS finds that the break in planet mass occurs at $7.7\pm1.7$\,M$_\oplus$ or  $q_{\rm br}=(2.5\pm0.6) \times 10^{-5}$ for the {\it Kepler} G-type median $M_{\rm host}$ of 0.91\,M$_\odot$. 
Within the uncertainties, $q_{\rm br}$ is the same as the one inferred in Section~\ref{sec:kepler} from the binned planet occurrence rate. The slope of the power law for $M < M_{\rm br}$ is shallower than the one reported from the binned data of MKG dwarfs, as EPOS better accounts for the incompleteness at small planet radii, while it is the same within the uncertainties for $M > M_{\rm br}$.


\begin{figure}[ht!]
\plotone{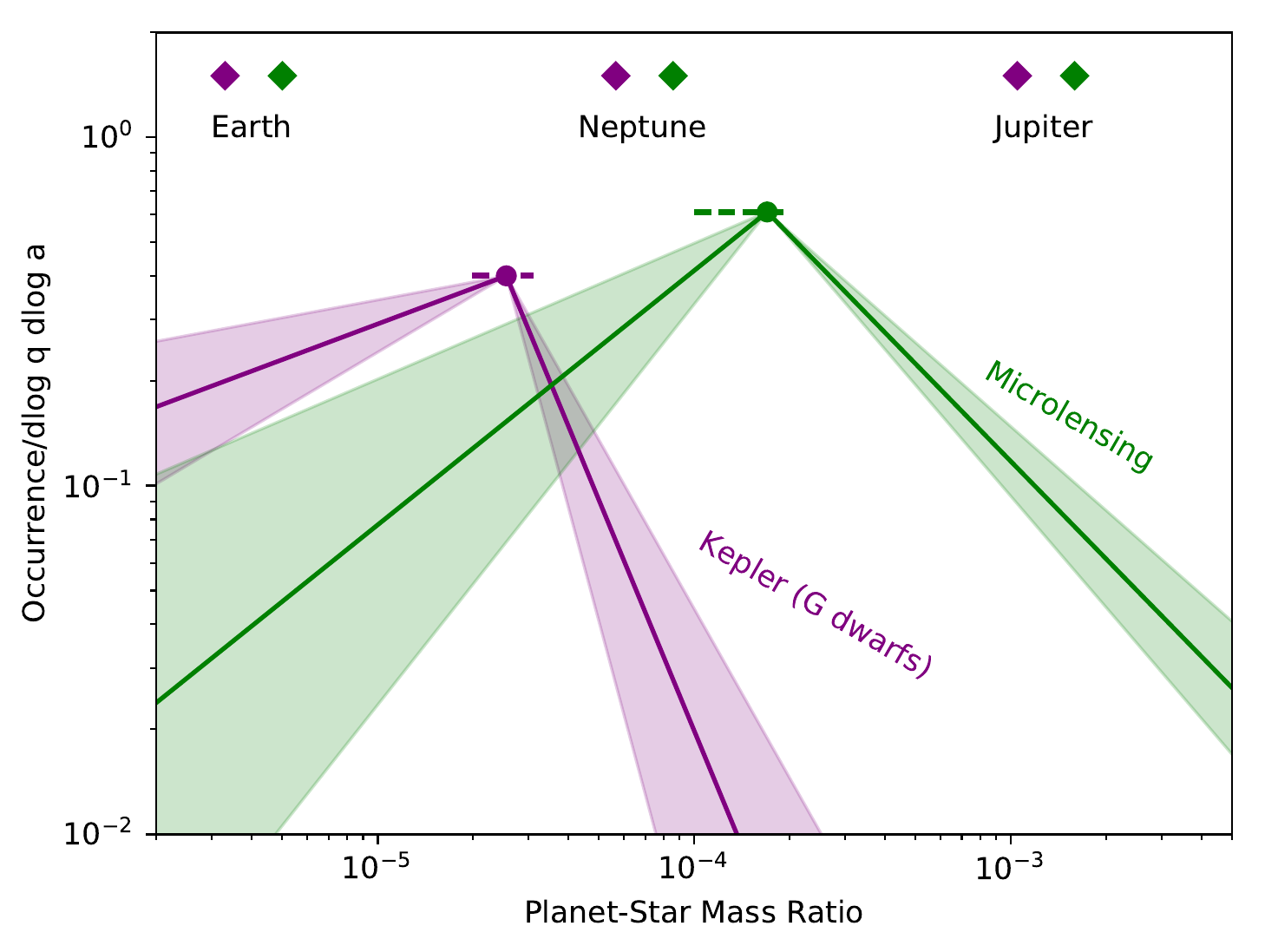}
\caption{Comparison of {\it Kepler} G dwarfs and microlensing results. The microlensing broken power law is from \citet{Suzuki2016} for a fixed $q_{\rm br}$ and $q > 1.7 \times 10^{-4}$ and from \cite{Udalski2018} for lower $q$. Colored diamonds at the top of the figure show the corresponding  $q$ for an Earth-, a Neptune-, and a Jupiter-mass planet using the median $M_{\rm host}$ of 0.91\,M$_\odot$ for {\it Kepler} G-type stars and 0.6\,M$_\odot$ for microlensing. \label{fig:epos}}
\end{figure}

Figure~\ref{fig:epos} compares the {\it Kepler} EPOS best fit broken power law to the one inferred via microlensing where the latter includes results from \citet{Suzuki2016} and \citet{Udalski2018}. In particular, the findings of  \citet{Udalski2018} place a lower limit on $q_{\rm br}$ of $10^{-4}$ as indicated in the figure.
Two properties of these functions are clearly distinct. First, the $q_{\rm br}$ inferred from the {\it Kepler} exoplanets is lower than that obtained from microlensing, as noticed earlier by \citet{Suzuki2016}. Considering our best fit parameters and uncertainties we can quantify that the difference is a  factor of  $\sim$3-10.  Second, the absolute value of the power law index for $q>q_{\rm br}$ is larger for {\it Kepler} than for microlensing. Taken together these two properties indicate that, around low-mass stars, the most common planet inside the snowline is smaller than the one outside and that there are fewer larger planets inside the snowline than outside. Figure~\ref{fig:epos} shows that  the occurrence at $q=10^{-4}$ is at least a factor of 7 lower for {\it Kepler} G dwarfs than for microlensing.
Considering that microlensing is mostly sensitive to planets around M hosts and that the {\it Kepler} normalization factor for M hosts is $\sim 2.5$ times higher than that for G hosts (see Table~\ref{tab:bestfit}), the difference in occurrence at  $q=10^{-4}$ 
reduces to a factor $\ga$3. This means that Neptune-mass planets around M dwarfs are at least three times less numerous inside than outside the snowline.

\section{Summary and Conclusions} \label{sec:discussion}
We computed the occurrence rate of exoplanets with radii $\sim1-6$\,R$_\oplus$ and periods $<$ 100 days vs planet-to-star mass ratio ($q$) and per stellar spectral type from the {\it Kepler} survey. Regardless of stellar host, the occurrence rate can be well described by a broken power law and we find that $q_{\rm br}$ and the power law indices are the same for MKG dwarfs, $M_{\rm host} < 1$\,M$_\odot$. Next, to better account for the survey incompleteness, we used EPOS on the G-type stars, as representative for the MKG dwarfs, and found that a 2D broken power law in orbital period and planet mass fit well the exoplanet population. In line with the results derived from the binned planet occurrence rates, the break in the power law occurs at $q_{\rm br} \sim 3\times 10^{-5}$. This value is $\sim$3-10 times lower than that identified by microlensing surveys whose planet hosts are mostly low-mass stars. These results are significant and timely in several ways.

First, they provide direct constraints to planet formation theories as they indicate that $q$ is a more fundamental quantity than planet radius or mass. The same mass-ratio function  means that the mass of the most common planet inside the snowline scales linearly with stellar mass for $M_{\rm host}<1$\,M$_\odot$. 
In the standard scenario of terrestrial planet formation,  the average planet mass  scales almost linearly with the local surface density of the planetesimal disk (e.g., \citealt{Kokubo2006}). A planetesimal surface density that scales linearly with stellar mass would then reproduce our results. However, dust masses of 1-3\,Myr-old disks follow a steeper than linear scaling relation with stellar masses, and the relation is even steeper in an older association (e.g., \citealt{Pascucci2016}). Either planetesimals form in less than 1\,Myr, when disk masses might scale linearly with stellar mass, or, if they form at later times, inward migration of pebbles (millimeter-/centimeter-grains) re-distributed the amount of solids in the inner disk. The latter scenario, in combination with streaming instability triggered by the accumulation of inward migrating pebbles at the snowline, has been put forward to explain the TRAPPIST-1 planetary system \citep{Ormel2017}. 
To first order this scenario can also explain the {\it Kepler} mass-ratio function as the pebble isolation mass scales linearly with stellar mass and the cube of the disk aspect ratio, see Equation~(15) in \citet{Ormel2017}, which however is the same for radial distances in the disk at the same temperature \citep{MD2012}.


Second, our results help to place the Solar System into context. With rocky planets interior to gaseous planets, the Solar System broadly follows the combined mass-ratio function inferred from {\it Kepler} and microlensing.  However, the most common planet mass inside the snowline is well above that of Earth,  pointing to accretion of a substantial gas envelope (e.g., \citealt{LF2014}),  while the most common planet outside the snowline is less massive than the gas giants in the Solar System. Thus, the exoplanet population appears to have a less extreme radial distribution of planetary masses than the Solar System. This is in line with a more efficient re-distribution of  solids  from the outer to the inner disk (e.g., \citealt{Izidoro2015}).


Finally, our best fit broken power laws as a function of planet orbital period, mass, and stellar type are key inputs to exoplanet yield calculations for future missions. Within the next four years, {\it Gaia} is expected to discover of order 10,000 new giant planets around A through to M dwarfs, many of which will be close to and beyond the snowline (e.g., \citealt{Perryman2014,Sozzetti2014}). Combined with the growing number of lower-mass planets discovered via transit, radial velocity, and microlensing, it will be possible to test whether the shape of the mass-ratio function beyond the snowline is also universal.



\acknowledgments
The authors thank Savita Mathur and Daniel Huber for sharing some of their results on stellar properties in advance of publication. I.P. also thanks Bertram Bitsch for a useful discussion on pebble accretion models.  This paper includes data collected by the {\it Kepler} mission. Funding for the {\it Kepler} mission is provided by the NASA Science Mission directorate.This material is based upon work supported by the National Aeronautics and Space Administration under Agreement No. NNX15AD94G for the program Earths in Other Solar Systems. The results reported herein benefited from collaborations and/or information exchange within NASAs Nexus for Exoplanet System Science (NExSS) research coordination network sponsored by NASAs Science Mission Directorate.

%

\vspace{5mm}
\facilities{{\it Kepler}} 


\software{{\tt astropy} \citep{2013A&A...558A..33A}, {\tt corner} \citep{fm2016}, {\tt emcee} \citep{Foreman-Mackey2013}
          }



\appendix


\section{EPOS parametric fit}\label{app:epos}
\begin{figure}[ht!]
\plotone{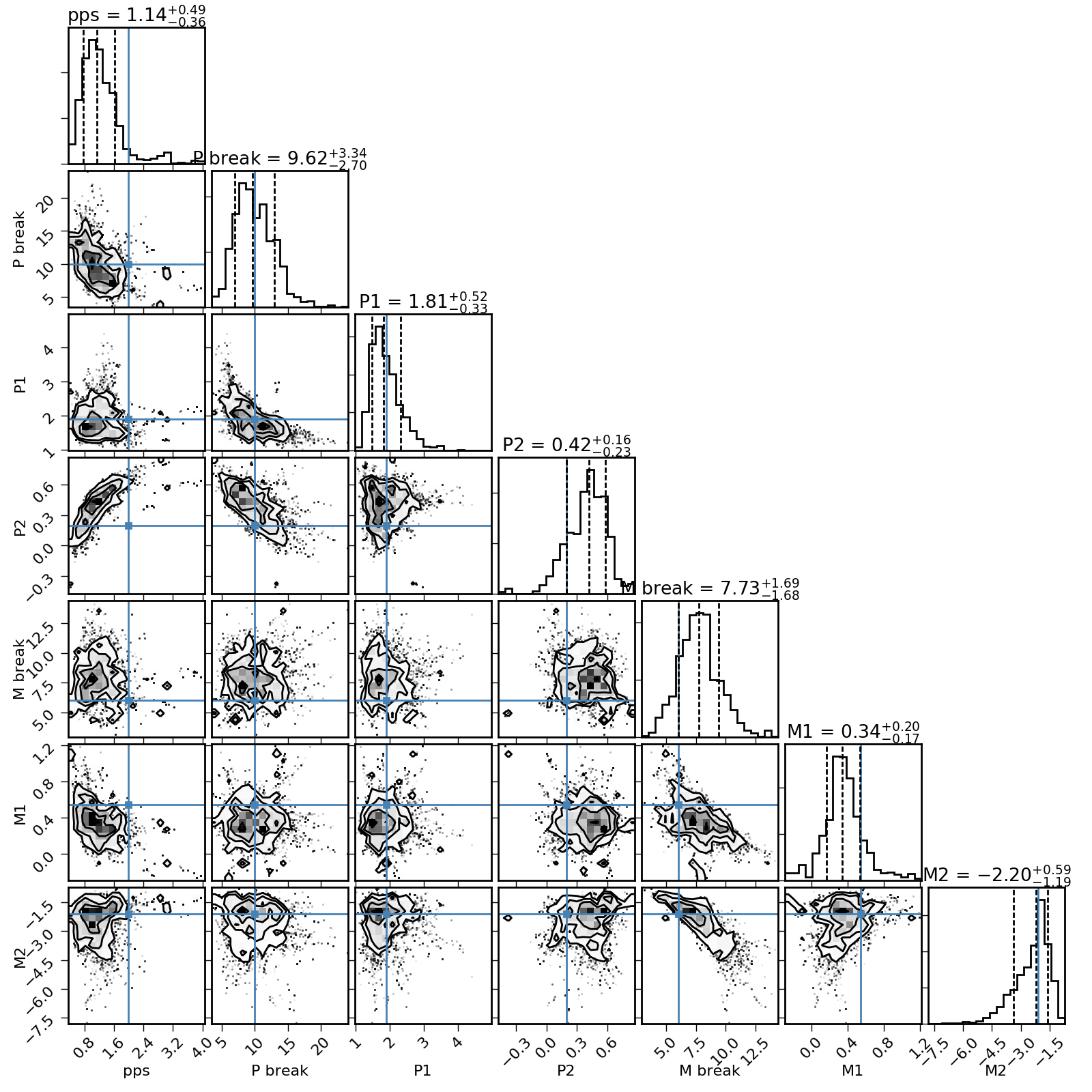}
\caption{EPOS corner plot showing the projections of the likelihood function for the 7 parameters that define the 2D broken power law occurrence rate vs orbital period and planet mass.  Blue lines indicates the initial guess. 
The corner plot was generated using the open-source {\tt Python} package {\tt corner} by \citet{fm2016}. \label{fig:triangle}}
\end{figure}

We use EPOS (Mulders et al. in prep.) in the occurrence rate mode to find the best parametric fit to the {\it Kepler} exoplanet population of G-type stars as representative for the MKG dwarfs, see Section~\ref{sec:kepler-moa}. A 2D broken power law in planet period ($P$) and radius is a standard function in EPOS as it is found to reproduce well the majority of the {\it Kepler} exoplanets (see Mulders et al. for details). For this project, the functional form is changed to a 2D broken power law in planet period and mass ($M$) of the following form:
\begin{equation}
{dN\over d\log P \, d\log M} = c_0 f(P) f(M) 
\label{eq:epos}
\end{equation}
where $c_0$ is a normalization factor and the integral of the function over the simulated planet period and mass range
equals the number of planets per star (pps). The planet orbital period distribution $f(P)$ is described as:
\begin{equation}
\begin{aligned}
f(P) =
\begin{cases}
\biggl( {P\over P_{\rm br}} \biggr)^{P_1} & \text{if $P \le P_{\rm br}$} \\
\biggl( {P\over P_{\rm br}} \biggr)^{P_2} & \text{if $P>P_{\rm br}$} 
\end{cases}
\end{aligned}  
\end{equation}
while the planet mass distribution $f(M)$ is described as:
\begin{equation}
\begin{aligned}
f(M) =
\begin{cases}
\biggl( {M\over M_{\rm br}} \biggr)^{M_1} & \text{if $M \le M_{\rm br}$} \\
\biggl( {M\over M_{\rm br}} \biggr)^{M_2} & \text{if $M>M_{\rm br}$} 
\end{cases}
\end{aligned}  
\end{equation}

\noindent where $P_{\rm br}$ and $M_{\rm br}$ are the breaks in orbital period and mass respectively. As discussed in Section~\ref{sec:kepler-moa}, when generating the synthetic populations the planet radii are calculated using a Monte Carlo calculation from the probabilistic mass-radius relation in  \citet{CK2017}.
Figure~\ref{fig:triangle} shows the best fit parameters and associated uncertainties with a run that used 100 walkers for 500 Monte Carlo iterations and a 200-step burn-in.  $M_1$ and $M_2$ are equivalent to $n$ in Equation~(1) for $q< q_{\rm br}$ and $q>q_{\rm br}$ respectively.

\end{document}